\documentclass[aps,prl,twocolumn]{revtex4}
\usepackage{graphicx}
\usepackage{epsfig}
\usepackage{amsmath}
\usepackage{amssymb}
\usepackage{amsfonts}%
\begin{document} 
\title{Magneto-spin Hall conductivity of a two-dimensional electron gas}
\author{M. Milletar\`i, R. Raimondi}
\affiliation{CNISM and Dipartimento  di Fisica "E. Amaldi", Universit\`a  Roma Tre, 00146 Roma, Italy}
\author{P. Schwab}
\address{Institut f\"ur Physik, Universit\"at Augsburg, 86135 Augsburg, Germany}

\begin{abstract} 
It is shown that the  interplay of long-range disorder and in-plane magnetic field 
gives rise to an out-of-plane spin polarization and a finite spin Hall conductivity of 
the two-dimensional electron gas in the presence of Rashba spin-orbit coupling. 
A key aspect is provided by the electric-field induced in-plane spin polarization. 
Our results are obtained first in the \textit{clean}  limit where the
spin-orbit splitting is much larger than the disorder broadening of
the energy levels  via the diagrammatic evaluation of
the Kubo-formula. Then the results are shown to hold in
the full range of the disorder  parameter
$\alpha p_F \tau$  by means of the quasiclassical Green function technique.
\end{abstract} 
\pacs{PACS numbers: } 

\date{\today} 
\maketitle 
It is  well established that the peculiar linear-in-momentum dependence of the Rashba 
(and of Dresselhaus) spin-orbit coupling leads in a two-dimensional electron gas (2DEG) 
to the vanishing of the spin Hall conductivity\cite{inoue2004,mishchenko2004,raimondi2005,khaetskii2006}.
This can be directly recognized by considering the continuity-like equation for the in-plane spin polarization, 
where the spin-nonconserving terms can be written as  the spin current associated to the out-of-plane 
spin polarization and to the spin Hall effect\cite{rashba2004,dimitrova2005,chalaev2005}.
In this paper, we show
that  the interplay of an in-plane magnetic field, $\mathbf{ \mathcal{B}}$, 
taken parallel to the electric field, $\mathbf{ \mathcal{E}}$, (say along the $\hat{\mathbf{ e}}_x$ axis) 
and long-range disorder changes this behavior providing then a potential handle on the spin Hall effect. 
In particular, we show that while the out-of-plane spin polarization
is linear in the magnetic field, 
the spin Hall conductivity is quadratic. 
Our analysis is valid  in the standard good metallic regime $\epsilon_F \tau /\hbar \gg 1$  
with spin-orbit effects  taken into account to first order in $\alpha /v_F$.
Here $\alpha$ and $\tau$ are the spin-orbit coupling  and the elastic quasiparticle lifetime due to impurity scattering, 
respectively, while $v_F$, $p_F$ and $\epsilon_F =v_F p_F /2$ are the parameters of the 2DEG in the 
absence of spin-orbit coupling.

Our proposal of a magnetic field-induced spin Hall effect  
differs from related previous suggestions both for the analytical treatment of it\cite{lin2006} 
and for the microscopic  mechanism responsible of the effect\cite{engel2007}. 
The difference of our proposal with respect to Ref.\cite{engel2007} is
closely related to the electric field-induced 
in-plane spin polarization, which for short-range disorder scattering is given by \cite{edelstein1990}
\begin{equation}
\label{edelstein}
s_y=-N_0 \alpha |e|\mathcal{E}_x\tau.
\end{equation}
In Eq.(\ref{edelstein}) $N_0=m/(2\pi)$ is the free density of states of the 2DEG in the absence of spin-orbit interaction. 
Contrary to what one could expect, the generalization of
Eq.(\ref{edelstein}) for long-range disorder, 
as we will show later, is not the replacement of the elastic quasiparticle lifetime $\tau$ with the transport time, 
$\tau_{tr}$, as it has been assumed in Ref.[\onlinecite{engel2007}]. 
We find then that long-range disorder leads to a non-trivial modification of the effective Bloch equations 
and eventually yields out-of-plane spin polarization and spin Hall effect. 
In contrast to Ref.[\onlinecite{engel2007}] we do not have to assume a
non-parabolicity of the energy bands or an energy dependence of the
scattering probability.

Our analysis is carried out in two steps. 
In the first step, we calculate the out-of-plane spin polarization
using the diagrammatic approach of 
Ref.[\onlinecite{raimondi2005}], 
valid in the \textit{clean} limit when the spin-orbit splitting 
is much larger than the disorder-induced broadening,  $2\alpha p_F \gg \tau^{-1}$.
In order to make contact with the  analysis  of Ref.[\onlinecite{engel2007}]  
performed in the opposite \textit{dirty}  limit ($\alpha p_F \ll \tau^{-1}$), we present, in the second step, 
a derivation based on the Eilenberger equation for the quasiclassical Green function 
in the presence of spin-orbit coupling\cite{raimondi2006}. 
The advantage of so doing is that the analysis is valid for an arbitrary value of the parameter $ 2 \alpha p_F \tau$
and also allows to determine the effective Bloch equations for the spin
density.

The Hamiltonian of a 2DEG perpendicular to the $\hat{\mathbf{ e}}_z$-axis reads
\begin{equation}
\label{eq1}
H=\frac{{\bf p}^2}{2m}+\mathbf{ b}(\mathbf{ p})\cdot \boldsymbol{\sigma}+V(\mathbf{ x})
\end{equation}
where ${\bf b}({\bf p})=\alpha\mathbf{ p}\times \hat{\mathbf{ e}}_z-\omega_s\hat{\mathbf{ e}}_x$ is
 the effective magnetic field  including both the Rashba spin-orbit coupling and 
the external magnetic field.
In Eq.(\ref{eq1}), $V(\mathbf{ x})$  describes the potential scattering from the impurities and, according to
 the established procedure in the literature,  will be taken
 as a random variable. 

The possibility of a non-vanishing spin Hall conductivity in the presence of an in-plane field 
may be appreciated by considering the equation of motion for the
$\hat{\mathbf{ e}}_y$-axis (in-plane) spin polarization, which yields
\begin{equation}
\label{eq3}
\frac{\partial s_y}{\partial t}+\frac{\partial }{\partial \mathbf{ x}}\cdot \mathbf{ j}^y_{s}
=-2m\alpha j_{s,y}^z+2\omega_s s_z,
\end{equation}
where $ j_{s,\gamma}^i$ is the $\hat{\mathbf{ e}}_\gamma$-axis component of the 
spin current that is polarized along the 
$\hat{\mathbf{ e}}_i$-axis. 
Under stationary and uniform conditions, the above equation implies, in the absence of the magnetic field, 
a vanishing spin current and hence a vanishing spin Hall conductivity, 
with the latter defined by $j^z_{s,y} =\sigma_{s,H} \mathcal{E}_x $.
In the presence of an in-plane  magnetic field one may have a bulk spin Hall current determined by
\begin{equation}
\label{eq5}
j^z_{s,y}=\frac{\omega_s}{\alpha m}s_z.
\end{equation}
In the following we evaluate $s_z$ to first order in the magnetic field which implies that  of $j^z_{s,y}$ to second order. 

\begin{figure}
\begin{center}
\includegraphics[width=3in]{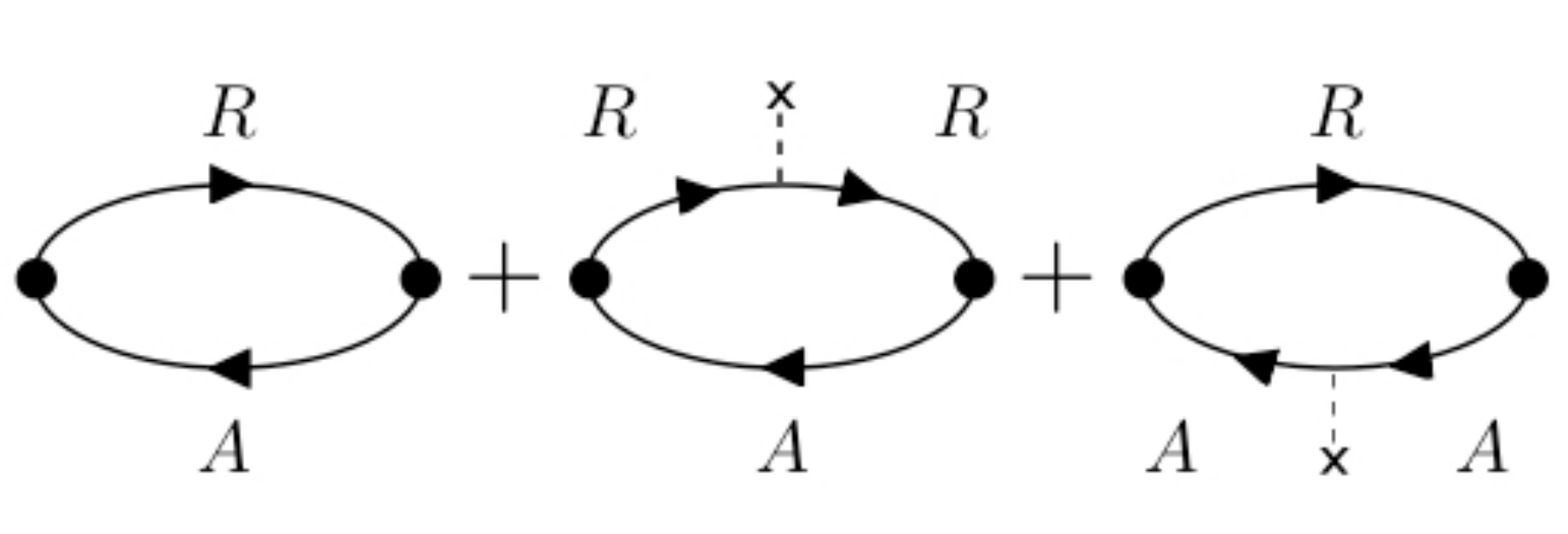}
\caption{Diagrams to be evaluated in zeroth  and first order in the
magnetic field. 
The dashed line ending with a cross indicates the magnetic field insertion. Under impurity average 
the (spin density and charge current) vertices are dressed by the standard ladder resummation.}
\label{kubo}
\end{center}
\end{figure}

As anticipated, we begin by sketching the  calculation performed with the diagrammatic approach. 
To linear order in the electric field, 
the Kubo formula for the zero-temperature expression of the out-of-plane spin polarization reads
\begin{equation}
\label{eq6}
s_z=-\frac{1}{2\pi}\sum_{\mathbf{ p}} \mathrm{Tr}_{\sigma}\left[
\overline{s_z G^R(\mathbf{ p})j_c^xG^A(\mathbf{ p})}
\right]|e|\mathcal{E}_x,
\end{equation}
where  the Green functions can be obtained from  Eq.(\ref{eq1}),  
the vertices are $s_z=(1/2)\sigma_z$,  $j_c^x=(p_x /m)\sigma_0 -\alpha \sigma_y$, 
and  the trace is over the associated spin indices.
In Eq.(\ref{eq6}), the bar indicates the average over the impurity potential, 
which, at the  level of the self-consistent Born approximation, yields  the self-energy 
\begin{equation}
\label{eq7}
\Sigma^{R,A}(\mathbf{ p})=\sum_{\mathbf{ p}'}  |V ({\mathbf{ p}-\mathbf{ p}'})|^2 G^{R,A}(\mathbf{ p}'),
\end{equation}
 $ |V (\mathbf{ q})|^2 $ being the Fourier transform of $\overline{V(\mathbf{x}) V(\mathbf{x}')}$.
In order to consider the effect of long-range disorder, we  expand the above scattering probability  as
\begin{equation}
\label{eq14}
 |V|^2=V_0 +2V_1 \cos (\varphi -\varphi ')+2V_2\cos (2\varphi -2\varphi')+\cdots ,
\end{equation}
where $\varphi- \varphi'$ is the angle between the two momenta
$\mathbf{p}$ and $\mathbf{p}'$.  
The harmonics, $V_0$, $V_1$, $V_2$ are functions of $|\mathbf{ p}|$
and $|\mathbf{ p}'|$.  In the following we will ignore this
dependence\cite{notefour} and take both momenta at $p_F$.
To first order in the magnetic field the diagrams to be evaluated are shown in Fig.\ref{kubo}. 
Notice that all the vertices and propagators appearing in the diagrams must be evaluated at zero magnetic field.
It is then convenient, following Ref.[\onlinecite{raimondi2005}], to use the disorder-free Hamiltonian eigenstates
\begin{equation}
\label{eq8}
|\mathbf{ p}\pm\rangle = \frac{1}{\sqrt{2}}
   \left(\pm \mathrm{i}\exp (-\mathrm{i}\varphi)|\mathbf{ p}\uparrow\rangle+|\mathbf{ p}\downarrow\rangle\right)
\end{equation}
corresponding to the eigenvalues $E_{\pm}=p^2/2m\pm \alpha p$ with $\tan (\varphi )=p_y /p_x$. 
In terms of the transformation matrix, $U$,  defined by 
Eq.(\ref{eq8}), the Pauli matrices transform as
\begin{equation}
\label{eq9}
\begin{array}{ccc}
U\sigma_x U^{\dag}&=&\hat{p}_y\sigma_z +\hat{p}_x\sigma_y \\
U\sigma_y U^{\dag}&=&-\hat{p}_x\sigma_z +\hat{p}_y\sigma_y \\
U\sigma_z U^{\dag}&=&-\sigma_x 
\end{array}, 
\begin{array}{ccc}
\hat{p}_x &\equiv &\cos (\varphi )\\
 \hat{p}_y&\equiv &\sin (\varphi )
\end{array}.
\end{equation}
As a consequence the spin density vertex, the magnetic field insertion and the charge current vertex become
\begin{eqnarray}
U s_z U^{\dag} & = & -\frac{1}{2}\sigma_x ,\label{eq10}\\
U (-\omega_s  \sigma_x) U^{\dag}  & = & -\omega_s (\hat{p}_y\sigma_z +\hat{p}_x\sigma_y), \label{eq11}\\
 U j_c^x U^\dag      &=&\left(\frac{p}{m}\sigma_0+\alpha \sigma_z\right)\hat{p}_x-\sigma_y\hat{p}_y.\label{eq12}
\end{eqnarray}
Upon impurity averaging the spin and charge vertices
get renormalized. In terms of the renormalized
quantities, Eq.(\ref{eq6}) becomes to first order in the Zeeman field
\begin{equation}
\label{eq13}
 s_z= - \mathrm{i}\omega_s\frac{|e|\mathcal{E}_x}{4\pi}\sum_{\mathbf{ p}}^{\mu=\pm}\hat{p}_x \mu
(\Gamma_{\mu\bar{\mu}} G^R_{\bar{\mu}}-\Gamma_{\bar{\mu}\mu}G^A_{\bar{\mu}})
G^R_{\mu}G^A_{\mu}J_{c,\mu\mu},
\end{equation}
where the first (second) term in the brakets refers to the magnetic 
field insertion in the top (bottom) Green function line and $\mu=\pm$ labels the eigenstates. 
The quantities  $\Gamma_{\mu \mu'}$ and $J_{c,\mu\mu'}$ are the dressed vertices 
corresponding to $s_z$ and $j_c^x$, respectively, 
and the Green functions are evaluated via the self-energy given in Eq.(\ref{eq7}). 
Apart from the spin vertex $\Gamma_{\mu \mu'}$, 
all the other quantities have been evaluated in Ref.[\onlinecite{raimondi2005}], 
where it has been shown that
 the off-diagonal  matrix element $J_{c,\mu\bar{\mu}}$
vanishes and the self-energy is diagonal in the eigenstate basis 
with 
\begin{equation}
\label{eq15}
\Sigma^{R(A)}_{\pm}= \frac{-(+)\mathrm{i}}{2\tau_{\pm}}, \ 
\tau_{\pm}=\tau \left( 1\pm \frac{V_1}{V_0}\frac{\alpha}{v_F}\right), \frac{1}{\tau}=2\pi N_0 V_0
.\end{equation}
The spin vertex obeys the equation
\begin{equation}
\label{eq16}
\Gamma_{\mu\bar{\mu}} =1+\sum_{\mathbf{ p'}}^{ \nu =\pm \mu}\Gamma_{\nu\bar{\nu}}
G^R_{\bar{\nu}}G^A_{\nu} |V|^2\frac{1+\mu \nu\cos (\varphi -\varphi ')}{2},
\end{equation}
which  yields 
\begin{equation}
\label{eq17}
\Gamma_{\mu\bar{\mu}} =1+ \frac{ \mathrm{i} \mu}{2 \alpha p_F \tau }
\frac{V_1}{V_0}.
\end{equation}
By using $G^R_{\mu}G^A_{\mu}=\mathrm{i}\tau_{\mu}(G^R_{\mu}
-G^A_{\mu})$,  integrating over the energy, $\xi=p^2/2m-\mu$, and keeping terms up to order 
$\alpha / v_F$,  Eq.(\ref{eq13}) becomes
\begin{equation}
\label{eq18}
 s_z
=-|e|\mathcal{E}_x\frac{ \omega_s}{8\alpha^2 \tau}
\left(1- \frac{V_1}{V_0} \right) \sum_{\mu =\pm}\mu  
\frac{N_{\mu} J_{\mu} \tau_{\mu}}{p_{\mu}^2},
\end{equation}
where $N_{\pm}=N_0(1\mp \alpha /v_F)$, $p_{\pm}=p_F(1\mp\alpha /v_F)$
are the density of states and the Fermi momentum of the two spin subbands 
and $ J_{\pm\pm}(p_{\pm})=J_{\pm}\hat{p}_x$ are the Fermi-surface expressions of the charge vertices.
 Finally, borrowing from Ref.[\onlinecite{raimondi2005}] the expression for the vertices
 \begin{equation}
\label{eq19}
J_{\pm}=v_F\left( \frac{V_0}{ V_0 -V_1}\mp\frac{\alpha}{v_F}
\frac{V_0 + V_2}{V_0-V_2}\right),
\end{equation}
one gets the out-of-plane spin polarization
 \begin{equation}
\label{eq20}
  s_z = -\frac{1}{2}|e|\mathcal{E}_x\frac{ \omega_s}{\alpha
  p_F}\frac{N_0}{p_F}\frac{V_1-V_2}{V_0-V_2}
\end{equation}
and, via  Eq.(\ref{eq5}), the spin Hall conductivity
\begin{equation}
\label{eq21}
\sigma_{sH}=-\frac{|e|}{4\pi}\left( \frac{\omega_s}{\alpha p_F}
\right)^2\frac{V_1-V_2}{V_0-V_2}.
\end{equation}
Remarkably the above central result shows that the sign of the spin Hall conductivity depends
on the relative strength of the harmonics of the scattering
probability.  This may explain the sign change
in the numerical evaluation of Ref.\cite{lin2006}. 
On the other hand, Eq.(\ref{eq20}) is inconsistent with Ref.\cite{engel2007}, 
where for the disorder model of Eq.(\ref{eq14}) one
would expect a vanishing out-of-plane polarization.

Recall that we derived Eqs.(\ref{eq20}) and (\ref{eq21}) in the
\textit{clean} limit.
We now rederive  the above results 
by means of the kinetic equations approach of
Ref.[\onlinecite{raimondi2006}] and find that Eqs.(\ref{eq20}) and
(\ref{eq21}) are valid for all values of the disorder parameter $\alpha
p_F \tau$. Furthermore we will derive the effective Bloch equations in
the \textit{dirty} limit.
We start with the Eilenberger equation 
\begin{eqnarray}
  \partial_{t} \check g & =& 
  -\frac{1}{2}\sum_{\mu = \pm } \left\{
    \frac{\bf p_\mu}{m}  \right.
   +\left. \frac{\partial}{\partial \bf p}({\bf b}_\mu \cdot {\boldsymbol  \sigma}),
    \frac{\partial}{\partial {\bf x} }\check g_\mu \right\} 
 \nonumber \\[1mm]
&-& {\rm i }\sum_{\mu = \pm } [{\bf b}_\mu \cdot {\boldsymbol \sigma}, \check
g_\mu ] - {\rm i} \left[ \check \Sigma , \check g \right]
\label{qc1}
\end{eqnarray}
for the quasiclassical Green function ($\check{ g } \equiv \check{g}_{t_1t_2}(\mathbf{ \hat{p}};\mathbf{ x}) $)
\begin{equation}
\label{qc2}
\check{ g } = \frac{\rm i}{\pi} \int {\rm d } \xi \, \check G_{t_1t_2}(\mathbf{ p},\mathbf{ x}), \
\check G = 
\left( \begin{array}{cc} 
G^R & G \\
0   & G^A 
\end{array}
\right)
\end{equation} 
where $\check{G}_{t_1t_2}(\mathbf{ p},\mathbf{ x})$ is  the Wigner representation of the
Green function, which has both matrix structure in the  Keldysh  (denoted by the  check symbol) and   spin spaces.
$[  , ]$ and $\{  ,  \}$ indicate  commutator and anticommutator.
As for the diagrammatic approach, the index $\mu=\pm$ labels 
the two spin subbands ($\mathbf{ b}_{\pm}=\mathbf{b }(\mathbf{
p}_{\pm})$). In integrations like in Eq.(\ref{qc2}) the corresponding
poles in the Green functions 
yield the two-component decomposition of  the quasiclassical Green function
\begin{equation}
\label{qc3}
{\check g}_{\pm}=\frac{1}{2}  \Big\{ \frac{1}{2}(1\pm {\hat {\bf b}}_0 \cdot {\boldsymbol \sigma}),{\check g}\Big\}.
\end{equation}
The "0" subscript denotes evaluation at the Fermi surface in the absence of spin-orbit coupling. 
In the following we are going to use Eq.(\ref{qc1}) to first order in the parameter 
$|{\bf b}_0|/\epsilon_F $.
The connection to the physical observables is made by integrating over
the energy $\epsilon$, which is the  Fourier conjugated variable of the time difference
$t_1-t_2$. For instance, the out-of-plane spin density
is given by the angular average of the Keldysh component\cite{notethree}
\begin{equation}
\label{qc3b}
s_z=s_z^{eq}
-\frac{N_0}{8}\int \mathrm{d}\epsilon \langle\mathrm{Tr}(\sigma_z g)\rangle, \
\langle ...\rangle\equiv \int_0^{2\pi}\frac{\mathrm{d}\phi}{2\pi}... \ .
\end{equation} 

In order to solve, to  linear order in the electric field, 
the Keldysh component of the Eilenberger equation (\ref{qc1}),  
we use the minimal substitution $\partial_{\mathbf{ x}} g \rightarrow -|e|\mathcal{E}_x{\hat {\bf e}}_x \partial_{\epsilon}  g_{eq}$
where $g_{eq}=\tanh (\epsilon / 2T) (g^R_{eq}-g^A_{eq})$ with
$g^R_{eq}=-g^A_{eq}=1- \partial_{\xi}\mathbf{ b}_0 \cdot {\boldsymbol \sigma}$ is
 the equilibrium quasiclassical Green function. 
As in the diagrammatic treatment previously developed, we find it
convenient  to transform the equations to the eigenstate basis via
Eq.(\ref{eq9}). 
After expressing the quasiclassical Green function as a four-dimensional column vector
\begin{equation}
\label{qc6}
 \tilde g = U g U^{\dagger}=\tilde{g_0} \sigma_0 +\mathbf{ \tilde{g}}\cdot {\boldsymbol \sigma}\rightarrow \left( \begin{array}{ cccc}
   \tilde{ g_0}&\tilde{ g_3}&\tilde{g_1}&\tilde{  g_2 }    
\end{array}
\right)^{t},
\end{equation}
the Eilenberger equation (\ref{qc1}) can be then written as a linear system of four equations for the components of $\tilde{g}$
\begin{equation}
\label{qc7}
\partial_t \tilde g = -\frac{1}{\tau}
(M_0+M_1) \tilde{g}+\frac{1}{\tau} (1+N)\langle K  \tilde{g}\rangle +S_0+S_1, 
\end{equation}
where, 
\begin{eqnarray}
N&=&-\frac{\alpha}{v_F}\left(
\begin{array}{cccc}
    0  &1&0&0    \\
1      & 0&0&0 \\
0&0&0&0\\
0&0&0&0  
\end{array}
\right) \\
M_0 &=& 1 + \frac{V_1}{V_0}N +  2 \alpha p_F \tau  \left(
\begin{array}{cccc}
     0 & 0 & 0 &0   \\
     0 & 0 & 0 &0 \\
     0 & 0 & 0 &1\\
     0 & 0 &-1 &0  
\end{array}
\right) \\
M_1 &=  &  2 \omega_s \tau \left(
\begin{array}{cccc}
  0&0&0&0\\
  0&0& -\hat{p}_x&0\\
   \frac{\alpha}{v_F} \hat{p}_x &  \hat{p}_x&0&-\hat{p}_y    \\
    0&0& \hat{p}_y &0
\end{array}
\right).
\end{eqnarray}
In Eq.(\ref{qc7}), $\langle ...\rangle$ denotes angle integration over $\varphi '$ with
 the scattering kernel that can be expandend into angular harmonics as
\begin{equation}
\label{qc8}
K (\varphi -\varphi')=K^{(0)}+\cos  (\varphi -\varphi')K^{(a)}
+ \sin (\varphi -\varphi')K^{(b)}+\cdots
\end{equation}
each coefficient being itself a matrix
$$
K^{(0)}=
\left(
\begin{array}{cccc}
    1  &0  &0   &0 \\
    0  &\frac{V_1}{V_0} & 0 &0\\   
      0&0&1&0\\
      0&0&0&\frac{V_1}{V_0}
\end{array}
\right),
K^{(b)}=\frac{V_0-V_2}{V_0}
\left(
\begin{array}{cccc}
     0  &0&0   &0 \\
      0&0 & 0&1\\   
      0&0&0&0\\
      0&-1&0&0
\end{array}
\right)
$$
and
$$
K^{(a)}= \frac{1}{V_0}
\left(
\begin{array}{cccc}
    2 V_1  &0  &0   &0 \\
          0&V_0+V_2 & 0&0\\   
      0&0&2V_1&0\\
      0&0&0&V_0+V_2
\end{array}
\right).
$$
Finally the \textit{source} electric-field dependent terms are
$$
S_0=E\left( 
\begin{array}{c}
      \hat{p}_x    \\
  -  \frac{\alpha}{v_F}  \hat{p}_x\\
    0 \\
   -  \frac{\alpha}{v_F} \hat{p}_y   
\end{array}
\right),
S_1= \frac{\omega_s}{v_F p_F} E\left(
\begin{array}{c}
0\\
      \hat{p}_x \hat{p}_y  \\
      0\\
      \hat{p}_x^2 
\end{array}
\right),
$$
with $ E=|e| \mathcal{E}_xv_{F} \partial_{\epsilon}(2\tanh (\epsilon /2T ))$.
Notice  that, consistently with the accuracy we are working,
one may use for the charge density component the solution obtained in the absence of both spin-orbit coupling
and magnetic field
\begin{equation}
\label{qc13}
j_{c,x} \sim \langle\hat{p}_x
\tilde{g}_0^{(0)}\rangle=\frac{1}{2}\frac{V_0}{V_0-V_1} \tau E,
\end{equation}
where the characteristic transport time renormalization $\tau_{tr} =
\tau  V_0 /(V_0-V_1)$ appears. 
We seek  now a stationary solution of Eq.(\ref{qc7}) which is
evaluated in first order in the magnetic field 
$\tilde g = \tilde{g}^{(0)}+\tilde{g}^{(1)} + \dots $. We then get 
\begin{equation}
\label{qc9}
\langle \tilde{g}^{(1)}\rangle=(M_0-K^{(0)})^{-1}( \tau \langle S_1\rangle-\langle M_1\tilde{g}^{(0)}\rangle).
\end{equation}
According to the transformation of Eq.(\ref{eq10}), the out-of-plane spin polarization is related to
\begin{equation}
\label{qc10}
s_z \sim \langle \tilde{g}^{(1)}_1\rangle=-
 \frac{\omega_s}{ 2\alpha p_F } \left( \frac{1}{v_F p_F} E
 + \frac{1}{  \alpha p_F \tau  }
 \frac{V_0 - V_1}{V_0 } \langle\hat{p}_x\tilde{g}^{(0)}_3\rangle
 \right),
\end{equation}
which is expressed in terms of $\langle\hat{p}_x\tilde{g}_3\rangle$ evaluated  at zero magnetic field.
This latter quantity is nothing but the in-plane spin polarization (cf. the second line of Eq.(\ref{eq9})). 
By  multiplying  the second component of the system Eq.(\ref{qc7}) by $\hat{p}_x$ and performing the angle average, one obtains 
the generalization, for long-range disorder, of the Edelstein result\cite{edelstein1990}
\begin{equation}
\label{qc12}
s_y \sim
\langle\hat{p}_x\tilde{g}^{(0)}_3\rangle=-\frac{\alpha}{v_F}\frac{V_0}{V_0-V_2}
\tau E.
\end{equation}
Finally, by using Eqs.(\ref{qc10},\ref{qc12}) into Eq.(\ref{qc3b}) 
one recovers the result (\ref{eq20}) of the diagrammatic approach,
which is now manifestly valid for any strength
of the disorder. 
To understand the meaning of Eq.(\ref{qc12}), 
it is useful to 
recall the origin of the in-plane polarization\cite{edelstein1990}: In the presence of an
electric field the Fermi surface is shifted by $\delta p_x \sim |e| \mathcal{E}_x \tau$. 
As a result the total spin of the
electrons neither in the plus nor in the  minus band adds up to
zero. Although the contributions of both bands tend to cancel, a finite
spin polarization remains due to the $\alpha/v_F$ corrections in the
density of states. For long-range disorder, the
Fermi surface shift is proportional to the transport time $\tau_{tr}$,
so one might expect the transport time also in the in-plane spin
polarization. However due to the $\alpha/v_F$ corrections 
each band has its own effective transport time, $\tau_{tr,\pm}\equiv
J_{\pm}\tau_{\pm}/v_F$ and 
the explicit result reads ($s_0 = \alpha N_0 |e| \mathcal{E}_x \tau$) 
\begin{equation}
\label{edelsteinrenormalized}
s_y=\frac{v_F}{4}(N_+\tau_{tr,+}-N_-\tau_{tr,-})|e|\mathcal{E}_x= -
\frac{V_0}{V_0-V_2} s_0. 
\end{equation}
At last 
we study the combined effect of magnetic and electric field in the 
 \textit{diffusive } regime, $\omega_s
\tau, \alpha p_F \tau \ll 1$.
The effective Bloch equations for the spin density are
\begin{eqnarray}
\partial_t s_x&=&-\tau_s^{-1}(s_x-s_x^{eq}) \label{qc14} \\
\partial_t s_y&=&-\tau_s^{-1} \left[ s_y +s_0 V_0/(V_0-V_2)^{-1} \right]+2\omega_s s_z \label{qc15}\\
\partial_t s_z&=&-2\tau_s^{-1} s_z-2\omega_s \left[s_y+s_0
\tau_{tr}/\tau  \right]\label{qc16},
\end{eqnarray}
where   $\tau_s^{-1}= 2 (\alpha p_F)^2 \tau_{tr} $ and $s_x^{eq}=N_0\omega_s$.  
Eqs.(\ref{qc14}) and (\ref{qc16}) agree with what was found in
Ref.\cite{engel2007}, the only difference is  
in the term proportional to the electric field (i.e. $s_0$) in Eq.(\ref{qc15}). 
The stationary spin polarization as a function of magnetic field is
now determined as 
\begin{eqnarray}
s_y&  =& -\frac{V_0 \, s_0 }{V_0-V_2}
\frac{1+2\omega_s^2\tau_s^2(V_0-V_2)(V_0-V_1)^{-1}}{1+2\omega_s^2\tau_s^2}\label{qc17} \\
s_z & =&  -\frac{V_0 \, s_0 }{V_0-V_2} \frac{V_1-V_2}{V_0-V_1}
\frac{\omega_s\tau_s}{1+2\omega_s^2\tau_s^2}\label{qc18}
\end{eqnarray}
showing an out-out-plane contribution as observed experimentally in
Ref.\cite{kato2004}.

In conclusion,  we have shown that the combined effect of an in-plane  magnetic field, long-range disorder and spin-orbit coupling gives rise
to an out-of-plane spin polarization and finite spin Hall
conductivity, whose value  does not depend on the concentration of
defects 
as long as the 2DEG is  in the metallic regime.
To obtain the correct value of the electric-field induced in-plane spin polarization
it is essential to take into account the different transport times in the two
spin-orbit 
splitted bands.

We acknowledge financial support by the Deutsche
Forschungsgemeinschaft through SFB 484 and SPP 1285 and by CNISM under  Progetti Innesco 2006.
R.R.  thanks the kind hospitality of the ICTS, Jacobs University, Bremen, where this work was initiated.


\begin{thebibliography}{999}
\bibitem{notethree} Within the quasiclassical formalism, $g$ describes the dynamic part only to which add the equilibrium part.
\end{thebibliography}
\end{document}